# BUBBLE GROWTH RATE IN SUPERHEATED DROPLETS

Tomoko Morlat[*]
Centro de Ciências e Tecnologias Nucleares, Instituto Superior Técnico,
Universidade de Lisboa, 2695-066 Bobadela, Portugal

**Abstract --** Calculations are presented to describe the dynamic of a growing bubble in a single and simple formulation for R(t). The calculations show clearly that the behavior of the growing bubble is exponentially increasing with the time constant τ showing a universal behavior as a function of the reduced superheat. Experimental data from other researchers are used to verify the description and are in good agreement with the results. The main motivation for this study was to find R(t)that is accurate and at the same time uncomplicated enough in solving the dynamic of the bubble growth to calculate the far-field pressure for superheated liquids.

## INTRODUCTION

Description of the bubble evolution R(t) following proto-bubble formation in superheated liquids has historically been divided between the early inertial stage when the growth is driven by inertial expansion and described by Rayleigh-Plesset (Rayleigh,1949; Plesset, 1949), and the later thermal stage governed by heat exchange and described by Plesset-Zwick (Plesset, 1954). In a recent work by Kozynets, Fallows and Krauss (Kozynets et al., 2019), a single description by Mikic (Mikic, 1970) was used to describe the full bubble evolution, permitting extraction of the acoustic pressure wave and simulation of the liquid response.

The Mikic description is however heuristic, and somewhat cumbersome. We here develop a single equation formulation for R(t) which shows an agreement with the experimental data reported by others. With R(t), the acoustic pressure is deduced as a function of time. The far-field pressure reproduces the signal amplitudes of reported results(Shepherd et al., 1981; Zhao, 2000).

The current state of the art is resumed in Sec. 1. Section 2 develops a single formulation for R(t) which is confronted with experimental data reported by Dergarabedian (Dergarabedian, 1953), Shepherd (Shepherd, 1981; Shepherd et al., 1982), Plesset-Zwick, and E. Soto (Soto et al., 2009). Section 3 describes the acoustic pressure and provides a second confirmation of the R(t) formulation.

1.  **BUBBLE EVOLUTION**

The description of a bubble evolution following nucleation is generally divided into two regimes: an early inertial stage when the rate is limited by strong liquid inertia, and a thermal stage when the temperature difference between the inside and outside of the bubble becomes large enough to initiate heat transfer.

In the first stage, the pressure of the liquid (p) can be extracted from the equation of motion

$$\frac{\partial \vec{v}}{\partial t} + (\vec{v}.\overrightarrow{grad}).\vec{v} = -\frac{\overrightarrow{gradp}}{\rho_l} \qquad (1)$$

,

where $v$ is the velocity of the liquid pushed outward, $\rho_l$ is the density of the liquid. After integration at the liquid-vapor interface,

$$\Delta P = \frac{3}{2} + \rho_l \left(\frac{dR}{dt}\right)^2 + \rho_l R \frac{d^2R}{dt^2} \qquad (2)$$

where $\Delta P = P_v - P_\infty$ where $P_v$ is the pressure of the bubble and $P_\infty$ is the operating pressure. The order of magnitude of the velocity which characterizes the inertial regime is such that the acceleration is $\rho_l R \frac{d^2R}{dt^2} \ll \frac{3}{2}\rho_l \left(\frac{dR}{dt}\right)^2$. Therefore, the solution is linear in time with $R_{inertial} = \sqrt{\frac{2\Delta P}{3\rho_l}} t$ : the bubble grows at a constant rate $\sqrt{\frac{2\Delta P}{3\rho_l}}$ at the expense of energy stored in the droplet itself. The inertial growth continues to push out the surrounding liquid ($\Delta T = T_\infty - T_v$ is still comparatively small so the growth continues linearly) until the $\Delta T$ becomes large enough to initiate heat transfer.

In the second stage, bubble growth becomes $R(t) \propto t^{1/2}$ (Plesset, 1954). The Jacob number $J_a = \frac{\rho_l c_{p,l} \Delta T}{\rho_v l}$, with $c_{p,l}$ the specific heat at constant pressure of the liquid, $\rho_v$ the vapor density and $\ell$ the latent heat of evaporation, describes the ratio between the available heat stored in the liquid and the latent heat required for vaporization. The larger is $J_a$, the less the latent heat is involved, the stored energy is larger than the latent heat and only a small part is converted into heat. Since Ja is proportional to $\Delta T$, one can say that the higher the superheat $\Delta T$ and the less the latent heat, the less is $\Delta T$ and the sooner the thermal regime starts. But at the same $\Delta T$, $J_a$ can be different – which is why the parameter $I_R = \frac{v_{in}^2}{v_{HT}^2}$ (ratio between inertial and heat transfer velocities squared) helps to understand which regime will dominate (Robinson, 2004; Lesage et al., 2014) :

$$I_R = \frac{v_i^2}{v_{HT}^2} = \left(\frac{4}{7}\right) \left(\frac{\gamma}{\rho_l D_l}\right) \frac{R_c}{Ja^2} \begin{cases} \ll 1 \to Inertial \\ \gg 1 \to Thermal \end{cases} , \qquad (3)$$

where $\gamma$ is the surface tension, $D_l$ is the thermal diffusivity of the liquid and $R_c$ is the initial bubble radius.

In between, both regimes are of the same order. An approximate solution is given by $R_{thermal} = A\sqrt{t}$ with $A = K^2 \frac{\rho_l c_{pl} \Delta T}{\rho_v l} \sqrt{D_l}$ and K depends on the specific approach and approximations made by various authors (Plesset, 1954; Alexsandrov, 1967; Prosperetti, 2017; Martynyuk et al., 1991; Peyrou, 1967; Brennen, 2005)(see Table 1). The crossover time ($t_c$) between the two stages is estimated by equating $v_{inertial}$ and $v_{thermal}$ (continuity of the velocities in changing regimes):

$$t_c = \frac{3}{8}\frac{\rho_l^3 c_{pl}^2 \Delta T^2 D_l}{\rho_v^2 l^2 \Delta P} = \frac{3}{2}K^2 Ja^2 \frac{\rho_l}{\Delta P}D_l \quad , \tag{4}$$

Table 1: Value of K according to different authors and their approach.

|   | Aleksandrov/ Prosperetti | Brennen | Peyrou | Plesset & Zwick | Martynyuk |
|---|---|---|---|---|---|
| $K$ | $2\sqrt{\dfrac{2}{\pi}} = 1.5958$ | $\sqrt{\dfrac{2}{3}} = 0.8165$ | $\sqrt{3} = 1.7321$ | $2\sqrt{\dfrac{3}{\pi}} = 1.9544$ | $2\sqrt{3\dfrac{T_{op}P_c}{T_c P_{op}}} = 17.32$ |

The approach of Plesset or Forster (Forster et al., 1954) considers a thin boundary layer in their description of the thermal regime. This assumption is correct as long as the Ja number is large enough. For the case of low Ja number or low superheat, Scriven (Scriven, 1959) obtained an expression which simplifies to $R(t) = \sqrt{2JaD_l t}$ knowing that in this case the energy available in the liquid is more or less the same order of magnitude as the latent heat. Therefore, the assumption of the thin layer is no longer valid for Ja < 2 and the curvature of the interface plays an important role. The model of Scriven incorporates the latent heat of phase change ($h_{lv}$) with specific heat ($c_{pl}$ and $c_{pv}$).

Table 2: Different formulations of the radius of thermal phase.

| Authors | Equation form |
|---|---|
| Aleksandrov/Prosperetti | $R = 2\sqrt{\dfrac{2}{\pi}}Ja\sqrt{D_l t}$ |
| Brennen | $R = \sqrt{\dfrac{2}{3}}Ja\sqrt{D_l t}$ |
| Peyrou | $R = \sqrt{3}Ja\sqrt{D_l t}$ |
| Plesset&Zwick | $R = 2\sqrt{\dfrac{3}{\pi}}Ja\sqrt{D_l t}$ |
| Martynyuk | $R = 2\sqrt{3\dfrac{T_{op}P_c}{T_c P_{op}}}Ja\sqrt{D_l t}$ |
| Scriven | $R = 2\sqrt{\dfrac{3}{\pi}\left(\dfrac{h_{lv}}{h_{lv}+(c_{pl}-c_{pv})\Delta T}\right)}Ja\sqrt{D_l t}$ |

These different approaches give an asymptotic solution for R(t) in the thermal stage.

The inertia-controlled bubble growth model ignores heat transfer, which is only valid in the early stage of bubble growth. In the thermal regime, on the other hand, the effect of inertia on bubble growth is neglected; this assumption is appropriate only during the late stage of bubble growth. A comprehensive bubble growth model must include both inertia and heat transfer, and provide a smooth transition between the two regimes.

A single formulation of R(t) covering the two growth regimes was made by Mikic and also Miyatake (Miyatake, 1994, 1997). Mikic used the Clausius-Clapeyron equation to evaluate the differential relation between saturated pressure and temperature by reexamining the combination of energy and momentum equations. Miyatake used a different approach considering a tiny bubble around the critical size. It is convenient to use them for calculation of bubble size at a specific growing time for a high Ja. But when $J_a < 10$, the model is insufficient to describe the behavior of the bubble growth because the solutions do not include the regime where the surface tension dominates.

Table 3: Single formulation of R(t).

| Authors | Equation form |
| --- | --- |
| Mikic et al. | $R^+ = \frac{2}{3}\left[(t^+ + 1)^{3/2} - (t^+)^{3/2} - 1\right]$ <br> $R^+ = RA/B^2$ <br> $t^+ = tA^2/B^2$ <br> $A = \sqrt{\frac{2h_{lv}\rho_v \Delta T}{3T_{sat}\rho_l}}$ <br> $B = \sqrt{\frac{12}{\pi}Ja^2 D_l}$ |
| Miyatake et al. | $R^+ = \frac{2}{3}\left[1 + \frac{t^+}{3}e^{-\sqrt{t^+ + 1}}\right]\left((t^+ + 1)^{3/2} - (t^+)^{3/2} - 1\right)$ <br> $R^+ = (R - R_c)\frac{A}{B^2}$ <br> $t^+ = \frac{A^2}{B^2}\left(t - t_d\left(1 - e^{-\left(\frac{t}{t_d}\right)^2}\right)\right)$ <br> $A = \sqrt{\frac{2\Delta P}{3\rho_l}}$ <br> $B = \sqrt{\frac{12}{\pi}Ja^2 D_l}$ <br> $R_c = \frac{2\gamma}{\Delta P}$ <br> $t_d = 6\frac{R_c}{A}$ |

## 2. REFORMULATION OF R(t) AND RESULTS

In this section, we develop a single formulation of R(t). The system will be called "droble" as it refers to the bubble (vapor) surrounded by the superheated droplet still remaining before total evaporation (see Fig. 1).

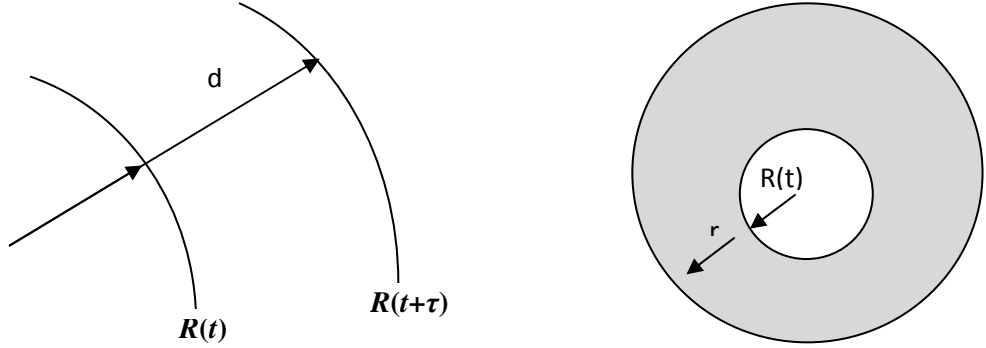

Fig. 1: (Left), schematic of the vapor radius expansion $R$ at time t and time t+τ. (Right), schematic of a droble with a vapor nucleus of vapor of size R(t) surrounded by the superheated liquid droplet. The position of a fluid element pushed by the vapor is represented at a distance r.

The viscosity is neglected since the ratio $\frac{\overrightarrow{gradP}}{\eta \Delta v} \gg 1$ (or equivalently $\eta \ll \frac{PR}{v} \sim \frac{10^5 \cdot 10^{-6}}{10} \Rightarrow \eta \ll 10^{-2} Pa.s$) implies low fluid viscosity of the liquids in this study (water and butane are of the order of $10^{-3}$Pa.s).

An intuitive approach is that the acceleration of the interface R(t) over a characteristic time τ (later called time constant) is equal to the velocity of the liquid layer pushed outward during the expansion, so that

$$\tau \frac{d^2R}{dt^2} = \frac{Dr}{Dt} = \frac{dr}{dt} \quad , \tag{5}$$

where Dr/Dt is a total derivative. We assume that the convective term $\vec{v} \cdot \overrightarrow{grad} v = 0$ since the exploration of the velocity field by the fluid particle will not explore areas of higher or lower velocity during its displacement.

Therefore:

$$\int \frac{d}{dt}\left(\frac{dR}{dt}\right) dt = \int \frac{dr}{\tau} \quad , \tag{6}$$

By integrating from $R$ to $R_{b,m}$, where $R_{b,m}$ is the final bubble size,

$$\frac{dR}{dt} = \frac{\Delta R}{\tau} = \frac{R_{b,m} - R}{\tau} \quad , \tag{7}$$

with solution is :

$$R(t) = R_{b,m}(1 - e^{-\frac{t}{\tau}}), \tag{8}$$

where τ is the time constant.

Another approach is if τ << t, *i.e.,* the case of the thin layer, where the expansion is rapid due to a high degree of superheat, in which case:

$$\vec{d} \cong \vec{R}(t + \tau) - \vec{R}(t) \tag{9}$$

Then $d \cong R(t) + \tau \frac{dR}{dt} - R(t)$ to first order, leading to $R_{b,m} - R(t) = \tau \frac{dR}{dt}$ or equivalently $\frac{dR}{dt} + \frac{R(t)}{\tau} = \frac{R_{b,m}}{\tau}$ giving the same solution as above.

Equation (8) matches all types of experimental data in Dergarabedian, Plesset-Zwick, Soto, and Shepherd for small/large $J_a$ and $I_R$. The time constant and the maximum size of the bubble can be extracted from the fit. For superheated droplets of known radius, $R_{b,m} = R_d \sqrt[3]{\frac{\rho_l}{\rho_v}}$ since mass conservation is assumed.

Various experimental results from others groups are represented below. The results of Soto are reproduced in Fig. 2 and represent the set of data fit with both Eq. (8) and Mikic models. It involves a single water drop in hot oil at 184°C with Ja = 355. The model gives a good fit. In Fig. 2(left), the data starts to show variation in time because the bubble started to oscillate after explosion, which the gap between the fit and the results after few msec.

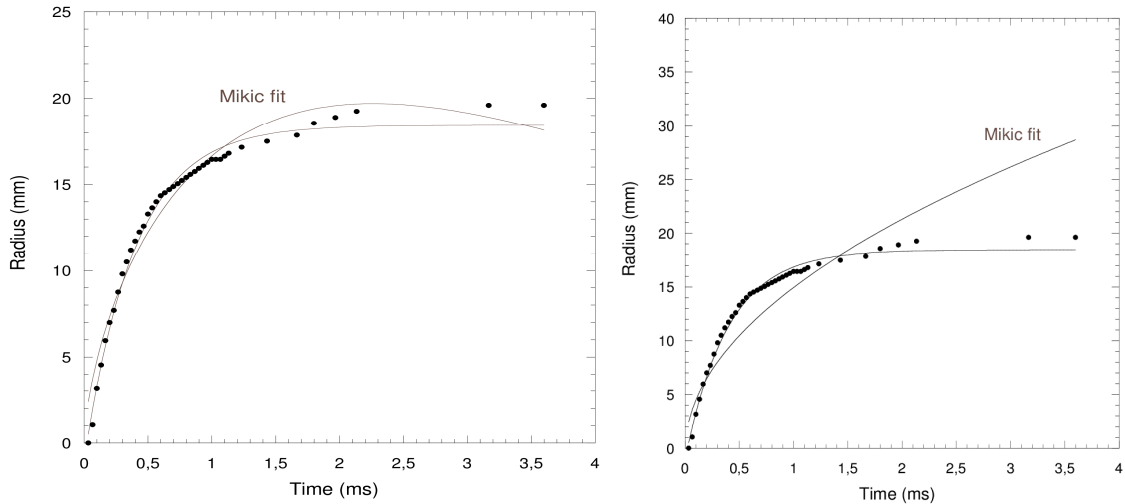

Fig. 2: Fit of both Mikic and model on Soto data set.

The results from Shepherd & Sturtevant (Shepherd et al., 1982)are reproduced in Fig. 3.With Ja = 317, the simple formulation of R(t) shows a good agreement between the data and the fit.

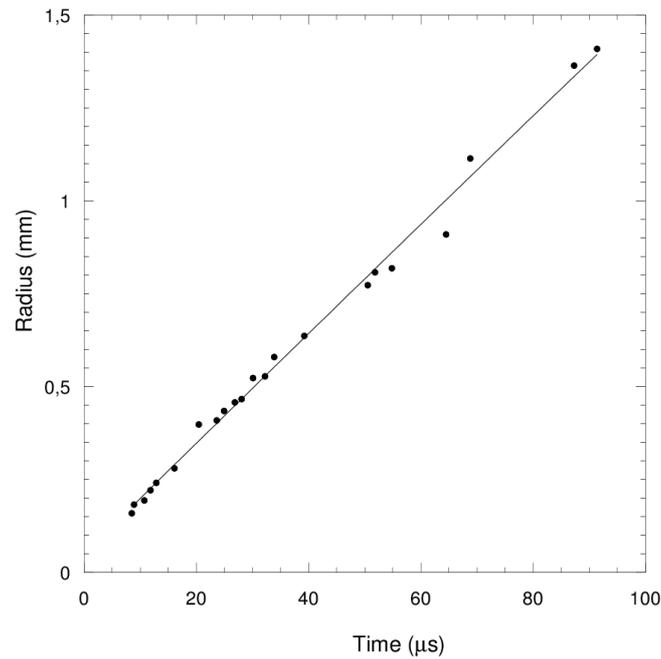

Fig. 3Shepherd, drop of butane at superheat limit with ΔT = 105.5°C and Ja = 316. The fit is in black line. The data are from Shepherd & Sturtevant.

In Fig. 4(right) below, the same experimental data is fit with Eq. (8). The data represent a drop of water in a low superheated state, with ΔT between 1- 5°C and low Jacob number: Ja = 3-15.The Plesset-Zwick model gives a R(t) ~ $t^{1/2}$ behavior close to experimental (Fig.4(left)); the Plesset-Zwick model over-estimates the results in the inertial stage, and under-estimates it in the thermal stage.

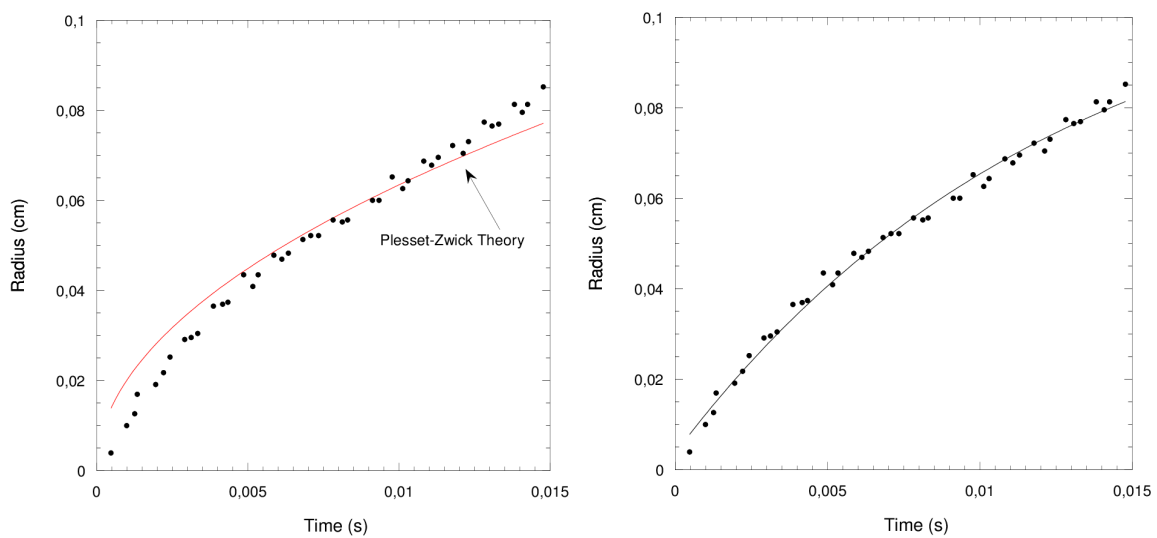

Fig. 4: On the left, the Plesset-Zwick fit (in red). On the right the exponential fit.

The experimental results of Dergarabedian, shown in Figs.5&6, describe the evolution of a water 'droble'. In Fig. 5, the data from Dergarabedian and Plesset-Zwick are superimposed since both used superheated water at a temperature of 104°C. Again, the fit of Eq. (8) matches all the data.

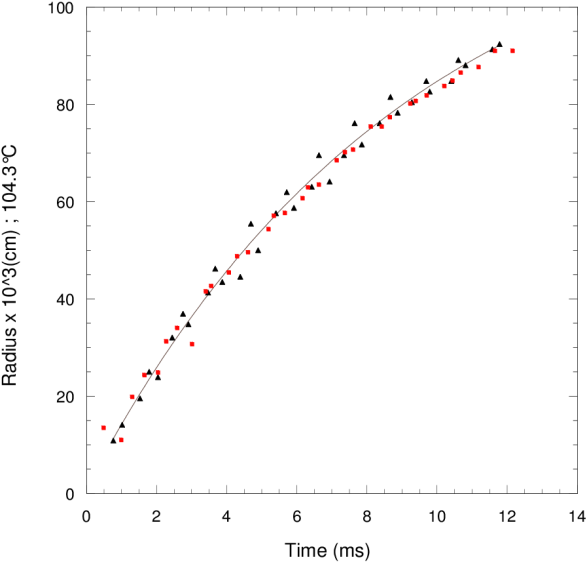

Fig. 5: Data from Dergarabedian (black triangles) and Plesset-Zwick (red squares) for water drop at 104.3°C. The fit is Eq. (8).

Figure 6 displays other experimental data from the Dergarabedian measurements for low Ja together with the expression of Eq. (8).

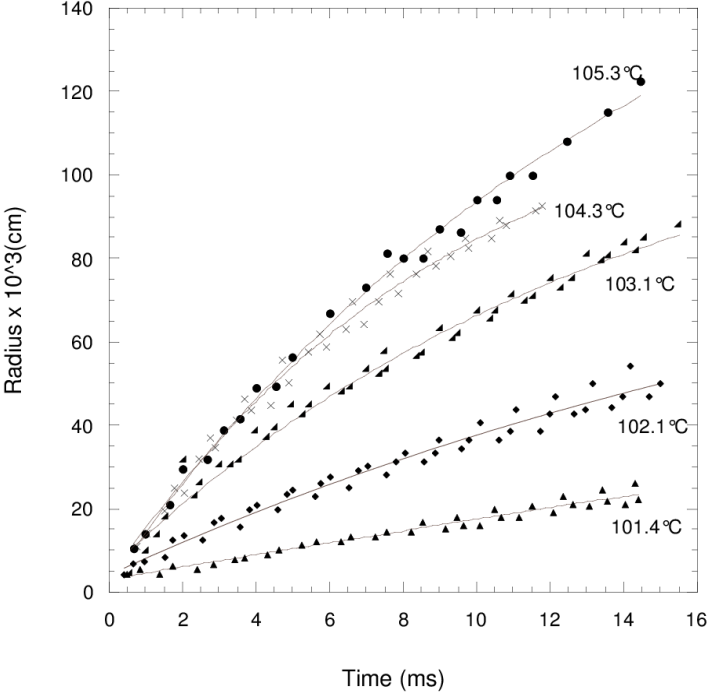

Fig. 6: All data from Dergarabedian represented with model.

Table 4 : Extraction of the time constant and final bubble size $R_{b,m}$ from the exponential fit. The reduced superheat factor is also displayed.

|  | Water |  |  |  |  |  | Butane |
|---|---|---|---|---|---|---|---|
| ΔT | 2.1°C | 3.1°C | 4.3°C | 4.5°C | 5.3°C | 84°C | 105.5°C |
| $\tau_{fit}$ | 25ms | 14ms | 9ms | 9ms | 12ms | 395µs | 112µs |
| $R_{b,m}$ | 1mm | 1.2mm | 1.3mm | 1.3mm | 1.6mm | 19.5mm | 8mm |
| Ja | 6.3 | 9.3 | 12.9 | 13.5 | 15 | 355 | 317 |
| $s = \dfrac{T - T_b}{T_c - T_b}$ | 0,0076642 | 0,011314 | 0,015693 | 0,016423 | 0,019343 | 0,30657 | 0,68950 |

Table 4 displays the values $R_{b,m}$ and $\tau_{fit}$, extracted from the exponential fit from Figure 2 to 6.

The data show that the higher the ΔT, the smaller the time constant τ. By using the definition of the crossover time given in Section1, for an average value of K = 1.6 such as $3/8K^2 = 1$, the crossover time $t_c$ for Soto and Shepherd are respectively 395µs and 155µs. Intuitively the time constant should be almost similar to the crossover time for high superheat degree. Indeed the Jacob number is large and therefore only a few percent of the energy stored in the droplet will be converted into latent heat (thermal phase); most of the energy is converted into kinematic. Therefore when the Ja number reaches large values (>>1), $t_c$ ~ τ. The expansion of the inertial phase is quickly reaching hundreds of microseconds. A thin layer of liquid remains at the end of the inertial regime and evaporates during the thermal phase. On the contrary when the degree of superheated is low, the expansion is slower, reaching few to hundred milliseconds. Therefore the thin layer hypothesis can no longer be used. A thick shell of superheated liquid surrounding the vapor nucleus remains, leading to a predominant thermal regime. Many authors have tried to describe the different regimes (mostly inertial or thermal or both depending on Ja and the ratio of inertial to thermal velocities) by different assumptions that match the data. The expression of Eq. (8) can be used for any Ja number.

The time constant can be calculated at high degree of superheat where $t_c$ ~ τ , by eliminating ΔP in the definition of $t_c$. The pressure is therefore dominated by the inertial regime and its definition is given by the surface tension ($p_s$) and hydrodynamic pressure ($p_{hd}$), with $p_\infty$ the operating pressure (far from the bubble):

$$p(t) = p_s(t) + p_{hd}(t) = p_\infty + \frac{2\gamma}{R(t)} + \rho_l R \frac{d^2 R}{dt^2} + \frac{3}{2}\rho_l (\frac{dR}{dt})^2 \qquad , \qquad (10)$$

By using the definition of Eq. (8) at t = τ,

$$p(\tau) = p_\infty + \frac{3.16\gamma}{R_{b,m}} - 0.03\rho_l \frac{R_{b,m}^2}{\tau^2} \quad , \tag{11}$$

and by using the definition of the crossover time for $t_c \sim \tau$ and $3/8K^2 = 1$ (Eq. (4)), one has $p(\tau) = Ja^2 D_l \frac{\rho_l}{\tau}$.

By equating the 2 pressures, one obtains an equation of second order in $\tau$ :

$$(p_\infty + \frac{3.16\gamma}{R_{b,m}})\tau^2 - D_l\rho_l Ja^2 \tau - 0.03\rho_l R_{b,m}^2 = 0 \quad , \tag{12}$$

yielding

$$\tau = \frac{D_l\rho_l Ja^2 \pm \sqrt{\Delta}}{2(p_\infty + \frac{3.16\gamma}{R_{b,m}})} \quad , \tag{13}$$

where $\quad \Delta = D_l^2 \rho_l^2 Ja^4 + 0.12\rho_l R_{b,m}^2 (p_\infty + \frac{3.16\gamma}{R_{b,m}}) \quad ,$ (14)

The calculation gives 451µs for Soto compared to $\tau_{fit}$ = 395 µs, and 112 µs for Shepherd compared to $\tau_{fit}$ = 112 µs.

## 3. ACOUSTIC PRESSURE

The purpose of this section is to validate the simple expression of R(t) in term of pressure change. The far-field pressure is directly calculated and confronted with experimental data. When a vaporization occurs, it emits an audible sound that can be recorded by a microphone.

The acoustic pressure amplitude is defined by

$$P_{ac}(t) = \frac{1}{4\pi d}\rho_\infty \frac{d^2 V(t)}{dt^2} = \frac{\rho_\infty}{d}\left(2R\left(\frac{dR}{dt}\right)^2 + R^2 \frac{d^2 R}{dt^2}\right) \quad , \tag{15}$$

where d is the distance of the nucleated droplet to the microphone and V is the volume of the droble in evolution. Substituting R(t) in Eq.(15) yields

$$P_{ac}(t) = \frac{\rho_\infty \rho_l}{\rho_v \tau^2 d} R_d^3 (1 - e^{-t/\tau})(3e^{-t/\tau} - 1)e^{-t/\tau} \quad , \tag{16}$$

Most papers relate the acoustic signal from a nucleation with oscillations. Applying Eq.(16) to the experimental data of Zhao (Zhao, 2000), we find good agreement as seen in the figure7. The acoustic pressure amplitude can be retrieved. This model doesn't consider the oscillations that frequently appear in the acoustic signal as seen in Fig.7 where the oscillations below zero correspond, according to the authors (Zhao, 2000), to maximum bubble volume, bubble shrinking and break up, bubble re-growth and maximum bubble for re-growth.

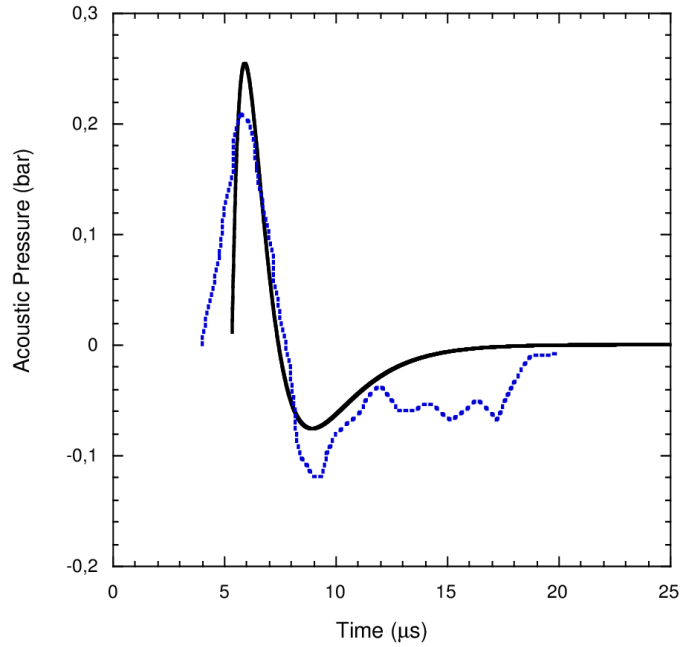

Fig. 7: Reproduction of the acoustic pressure from Zhao (blue dot line)and the acoustic pressure from the model (thick line).

The data from Shepherd (Shepherd et al., 1982) using a drop of butane at the superheated limit is shown in Fig.8. The time constant is first extracted from the reference of Shepherd to plot the resulting acoustic pressure.

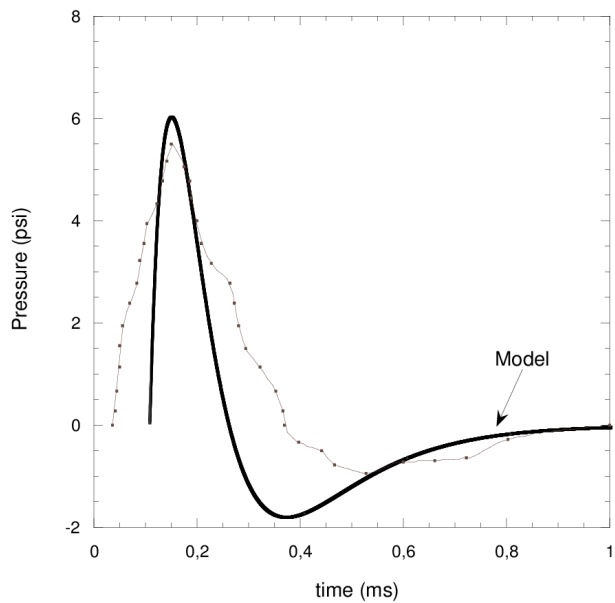

Fig. 8: Reproduction of the experimental acoustic signal from Shepherd with the model superimposed.

The model reproduces the acoustic pressure amplitude with τ = 112 µs. The experimental data from Shepherd as seen in Fig.8seems to be stretched over the abscissa. Some effects from the data acquisition such as amplification, distortion of the signal via dispersion of the sound propagation in the media suffice to explain the width of the experimental results, but the height of the acoustic pressure amplitude is retrieved by Eq. (16).

## 4- DISCUSSION

The expression for R(t) is accurate and uncomplicated in solving the dynamic of the bubble growth. The model is also capable of retrieving the shape of the acoustic pressure signal and also the amplitude of the signal. This description doesn't take into account the oscillations observed in the acoustic signal as referenced in Shepherd, Kwak or Park (Kwak, 1995; Park, 2005). Indeed according to these authors, oscillations occur after total evaporation of the liquid on a millisecond time scale (caused by an interfacial instability driven by rapid nucleation, suggested by the Landau mechanism as instability of laminar flames (Shepherd, 1982)).

Some studies (Wang, 2017) reference the long delay period of the initiating bubble growth before an obvious increase of its size as observed for low superheat or reduced pressure. In this case, the expression can be rewritten as $R(t) = R_{b,m}(1 - e^{-\frac{(t-t_0)}{\tau}})$ where $t_0$ is the time delay.

The time constant plays an important role in determining the growth rate of the bubble. In section 3, we demonstrate one way of calculating τ for high superheat degree (or high Ja number) by eliminating the internal pressure ΔP. By using the results of Table 4 for either water and butane, an intuitive way of determining the time constant at any Ja number would be to plot the time constant τ as a function of the reduced superheat factor $s = \frac{T-T_b}{T_c-T_b}$ where $T_b$ and $T_c$ are the boiling and the critical temperatures respectively (d'Errico, 2001). Then s = 0 at $T_b$, and 1 at $T_c$: s defines the degree of metastability of the system within the interval. When displayed as a function of s, the time constant τ lies on a respective "universal" curve as shown in Fig. 9. The time constant behaves as $\tau = k.s^q$ where k = 0.088526 and q = -1.154 (Fig. 9). Since the liquids are similar in their thermophysical properties, the "universality" of their properties is not particularly surprising. One can complete or improve the universality curve by using other liquids, as well as by varying the superheat. Therefore, the entire process of bubble growth can be deduced directly from the universal τ curve.

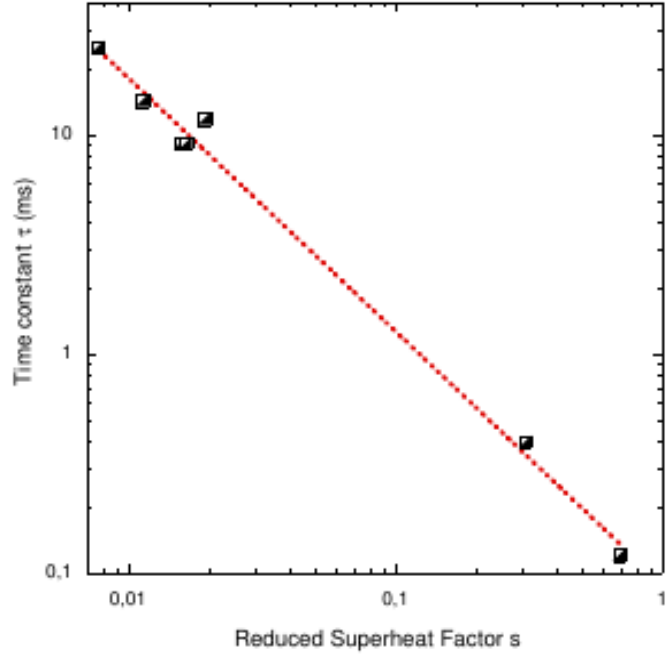

Fig. 9 : Time constant τ as a function of the reduced superheat factor s. The red fit is a power law.

Experiments using a microphone for acoustic detection can also extract the time constant from the acoustic signal. The value of the maximum amplitude is obtained from $\frac{dP_{ac}}{dt} = 0$, which gives $t_{risetime} = \tau \ln\left(\frac{9}{4+\sqrt{7}}\right)$ where $t_{risetime}$ is extracted from the experimental data. Therefore $\tau_{exp} = \frac{t_{risetime}}{\ln\left(\frac{9}{4+\sqrt{7}}\right)} = \frac{t_{risetime}}{0.3032}$. By retrieving the time constant and the final size of the bubble, one can define the entire process of the bubble growth.

ACKNOWLEDGEMENTS

I thank Tom Girard for useful discussions on the universality of metastable liquid properties and also for careful reading of this manuscript. This work was funded in part by the Portuguese Foundation for Science and Technology (FCT) via grant UID/Multi/04349/2013 and grant PTDC/EEI-ELC/2468/2014.

**References**

Alexsandrov, Yu.A., Voronov, G.S., Gorbunkov, V.M., Delone, N.B., Nediayov, Yu.I.BubbleChambers.(Indiana University Press, Bloomington and London, 1967).

Brennen, C.E. Bubble Growth and Collapse.Fundamentald of Multiphase Flow, Chp. 4 (Cambrigde Press University, 2005).

Dergarabedian,P. The Rate of Growth of Vapor Bubbles in Superheated Water. ASMF, J. of Appl. Mech. 20(1953)537-45.

d'Errico, F. Radiation Dosimetry and Spectrometry with Superheated Emulsions.Nucl. Instrum. Meth. B184 (2001) 229.

Forster,H. K., Zuber,N. Growth of a vapor in a superheated liquid. J. Appl. Phys. 25 (4) (1954)474-478.

Kozynets, Fallows and Krauss.Modeling emission of acoustic energy during bubble expansion in PICO bubble chambers. Phys. Rev. D 100 (2019)052001.

Kwak,H.-Y., Oh,S.-D., Park,C.-H. Bubble Dynamics on the Evolving Bubble Formed From the Droplet at the Limit of Superheat. Int. J. Heat Mass Transfer 38 (1995) 1709.

Lesage,F.J., Siedel,S., Cotton,J.S., Robinson,A.J. A mathematical model for predicting bubble growth for low Bond and Jakob number nucleate boiling. Chem. Eng. Science, 112 (2014) 35-4.

Martynyuk,Yu.N. and Smirnov,N.S. Sound Generation in Superheated Liquids by Heavy Charged Particles. Sov. Phys. Acoust. 37 (1991) 376.

Mikic,B.B., Rohsenow,W.M., Griffith,P. On Bubble Growth Rates. Int. J. Heat Mass Transfer 13 (1970) 657.

Miyatake,O., Tanaka,I., Lior,N. Bubble growth in superheated solutions with a non-volatile solute. Chem. Eng. Science, Vol. 49, No. 9(1994) 1301-1312.

Miyatake,O., Tanaka,I., Lior,N. A simple universal equation for bubble growth in pure liquids and binary solutions with a non-volatile solute.Int. J. Heat Mass Transfer Vol. 40, No. 7 (1997) 1577-1584.

Park, H-C., Byun, K-T, Kwak, H-Y. Explosive boiling of liquid droplets at their superheat limits. Chem. Eng. Science 60 (2005), 1809-1821.

Peyrou,Ch. Bubble Chamber Principles. Bubble and Spark Chambers, Vol. 1, Ed. R. P. Shutt (Academic Press, New York, 1967).

Plesset,M.S. The Dynamics of Cavitation Bubbles. ASME J. Appl. Mech. 16 (1949) 228.

Plesset,M.S. and Zwick,S.A.The Growth of Vapor Bubbles in Superheated Liquids.J. Appl. Phys.25 (1954) 493.

Prosperetti,A. Vapor Bubbles. Ann. Rev. Fluid Mech. 49 (2017) 221.

Rayleigh,L.On the Pressure Developed in a Liquid During the Collapse of a Spherical Cavity, Phil. Mag. 34 (1917)94;

Robinson,A.J., Judd, R.L. The dynamics of spherical bubble growth. Int. J. Heat Mass Transfer 47 (2004) 5101–5113.

Shepherd,J.E. Dynamics of Vapor Explosions. Rapid Evaporation and Instability of Butane Exploding at the Superheat Limit. PhD Thesis, California Institute of Technology, 1981, p. 159;

Shepherd,J.E. and Sturtevant,B. Rapid Evaporation at the Superheat Limit.J. Fluid Mech. 121 (1982) 379-402.

Soto,E., Zenit,R.and Belmonte,A. Superheated water drops in hot oil. arXiv:0910.3272v1 [physics.flu-dyn] 17 Oct 2009. Videos: https://ecommons.cornell.edu/handle/1813/14113.

Scriven,L. E. On the dynamics of phase growth. Chem. Eng. Science, 10 (1959) 1-13.

Wang,Q., Gu,J., Li,Z., Yao,W. Dynamic modeling of bubble growth in vapor-liquid phase change covering a wide range of superheats and pressures. Chem. Eng. Science 172 (2017)169-181.

Zhao,A., Glod,S., Poulikakos,D. Pressure and Power Generation during Explosive Vaporization on a Thin-film Microheater. Int. Journ. of Heat and Mass Transfer 43 (2000) 281.